\documentclass[twocolumn]{elsart}
\usepackage{epsfig}
\usepackage{graphics}
\begin{document}
\begin{frontmatter}
\title{High Rate Performance of Drift Tubes}
\author{G. Scherberger,}
\author{V. Paschhoff,}
\author{V. Waldmann,}
\author{U. Landgraf,}
\author{G. Herten and}
\author{W. Mohr}
\address{Fakult\"at f\"ur Physik der Universit\"at Freiburg,
Hermann-Herder-Str. 3, D-79104 Freiburg i.\,Br., Germany}

\begin{abstract}

This article describes calculations and measurements of space charge
effects due to high rate irradiation in high resolution drift
tubes. Two main items are studied: 
the reduction of the gas gain and changes of the drift time.
Whereas the gain reduction is similar for all gases and unavoidable, 
the drift time changes depend on the kind of gas that is used.
The loss in resolution due to high particle rate can be minimized with
a suitable gas. This behaviour is calculable, allowing predictions for
new gas mixtures.
\end{abstract}
\end{frontmatter}

\section{Introduction}

The influence of space charges in  proportional counters has been
studied  theoretically and experimentally and is described in various
articles \cite{hen69,paper2}.  There, the main emphasis
was put on the drop of the gas gain which is important for measuring
charges.\\
For drift chambers in which no charge measurement is foreseen,
the  gain drop is of secondary importance and  the drift time is the
relevant information. Variations in the drift time due to
a disturbed electric field lead to a loss in the spatial resolution.\\
This work was done in the context of the development of
the {\sc Atlas} muon 
spectrometer where high background rates coming from neutrons and
photons are expected. The detector should still work at background
rates of 500\,Hz/cm$^2$ (5 times the expected
background)\cite{tdr} -- with a spatial resolution for a single tube
better than 100\,$\mu$m. \\
The {\sc Atlas} muon spectrometer will be built from 3\,cm diameter drift
tubes with a 50\,$\mu$m wire in the middle.
The gas pressure is raised to 3\,bars absolute in order to reach the
desired spatial resolution at the nominal gas gain of  $2\cdot
10^4$. If not
stated otherwise, all the measurements and calculations  described
here refer to these operating conditions.\\
Important criteria for the choice of the gases are given by the need of a
non-flammable gas with a maximum drift time well below 1\,$\mu$s. 
Up to an accumulated charge deposition of 0.6\,C per cm wire, ageing
effects should be excluded.\\ 
The outline of this article is as follows: in section\,2 a calculation
of the space charge effects is shown, section\,3 gives a description of
the experimental setup in the test beam. Section 4 
shows how the data readout and analysis was done. Section\,5 gives
results of the gas gain reduction, followed by the treatment of the
changes in the drift time in section\,6. The event-to-event
fluctuations which are responsible for the irreducible loss of the
spatial resolution are described in section\,7.

\section{Calculation of Field Disturbances}
\label{calcsect}
The electric field inside a drift tube with radius $ b $ and
wire radius $ a 
$ held at potential $ V $ is given by:
\begin{equation} E(r) = \frac{V}{\ln{\frac{b}{a}}} \cdot \frac{1}{r}
\label{eq:ef} \end{equation}
This formula is only correct if one can neglect the positive ions
that are produced in the avalanche processes. They drift towards the
cathode and disturb the electric field. They screen
the positive potential at the wire and lead to a reduction of the
electric field near the wire. Thus one expects a lower gas gain.
Since the total voltage between wire and tube is kept constant, the electric
field at large radii is increased if the field near the wire is
decreased. One expects a change in the drift time because the drift
velocity for electrons depends on the electric field.\\
The density of the space charges coming from positive ions can be
calculated assuming a homogeneous irradiation within one tube, thus
neglecting all effects of event-to-event fluctuations.
These fluctuations will not change the mean value of the charge
density but affect the resolution as will be shown in the last
section. \\
The drift velocity $w$ of the positive ions is proportional to the
reduced electric field 
\begin{equation} w = \mu \cdot \frac{E}{p}\ \ ,\label{eq:mue}
\end{equation}   
where $p$ is the gas pressure,
$\mu$ the mobility of the gas and the electric
field is given by (\ref{eq:ef}). 
The maximum ion drift time $t_+$
(which is the drift time for almost all ions because nearly all of
them  are produced at
the wire) is obtained by integrating (\ref{eq:mue}),
\begin{equation} t_+  = \frac{b^2 p \ln{\frac{b}{a}}}{2 \mu V} \ \
.\label{eq:t+} 
\end{equation} 
Note that for the calculation of the above formula the electric field
(\ref{eq:ef}) of the  
undisturbed tube  is used. For high rates, the electric
field will change and the maximum ion drift time has to be corrected,
as will be shown below.\\
From  \cite{hen69}\footnote{the results shown there are
in error by a factor of 
$\pi$} one can see that the density of the ions $\rho$ is independent
of the radial 
distance $r$ from the tube centre 
in case of a cylindric tube geometry for a homogeneous irradiation.
Since $ w \propto 1/r $ the density of the ions is:
\begin{equation} \rho(r) = \rho = n_p \cdot g \cdot t_+ \frac{1}{\pi
b^2} \cdot \frac{R}{L} \label{eq:rho} 
\end{equation}
with $ n_p $  = the number of primary ion pairs (the number of pairs
before gas amplification), $ g $ is the gas amplification factor and 
$ R/L $  is the par\-ticle rate per wire length. 
The potential $\Phi(r)$ can now be calculated with Poisson's equation 
\footnotesize
\begin{equation} \triangle \Phi = - \frac{\rho e}{\varepsilon_0}
\hspace{3mm} \protect{\rm (mksa \ \ units)} \end{equation}
and the boundary conditions  $ \Phi(a) = V $ and $ \Phi(b)=0 $. 
\begin{eqnarray}
\Phi(r)  & =& \frac{V \ln{\frac{b}{r}}}{\ln{\frac{b}{a}}} 
          +  \frac{\rho e}{4 \varepsilon_0}  \nonumber \\
 & \times & 
 \left[ \left(b^2 -r^2 \right) - \frac{ \left(b^2-a^2\right)
\ln{\frac{b}{r}}}{\ln{\frac{b}{a}}} \right] 
\end{eqnarray}
\normalsize
Thus
\footnotesize
\begin{eqnarray}
 E(r) & = &  -\nabla \Phi(r) \label{eq:newfield} \\
      & = & \frac{V}{\ln{\frac{b}{a}}} \cdot
\frac{1}{r} + \frac{\rho e}{4 \varepsilon_0} \left[2 r -
\frac{b^2-a^2}{\ln{\frac{b}{a}}} \cdot \frac{1}{r} \right] \nonumber
 \end{eqnarray}
\normalsize
This formula has to be taken instead of (\ref{eq:ef}). \\
\begin{figure}[t]
\centerline{\psfig{figure=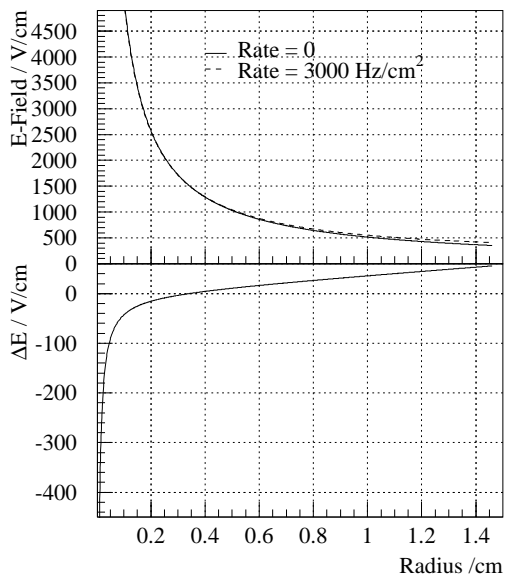,width=\columnwidth}}
\caption{\protect\footnotesize Comparison of the electric field inside
a tube with  and without space charges.
Top: electric fields, bottom: difference of
the electric fields.  The field near the wire which
is relevant for the gas amplification is decreased whereas the field
for the drifting electrons is increased for $ r >$ 0.4\,cm.  }
\label{fig=efield}
\end{figure}
Figure 
\ref{fig=efield} (top) shows an example of the electric field inside a tube with and
without the addition of space charges coming from a muon rate of 3000
Hz/cm$^2$. A 3\,cm diameter tube with a 50\,$\mu$m diameter wire is
operated at a 
gain of $2 \cdot 10^4$ at a pressure of 3\,bar with the gas
Ar-CH$_4$-N$_2$ 91-5-4\,\%. One can see that the
field for drifting electrons is changed. The difference plot (bottom)
shows clearly that the field is reduced in the region near the wire,
but increased for radial distance\,$>$\,4\,mm. 
As the drift velocity
depends on the electric field, the time to drift a certain distance
will be different: the r-t relation changes with rate. The magnitude
of this effect depends on the variation of the drift velocity with
E, which is strongly dependent on the gas.\\
For calculations of the gas gain, only the field near the wire has to
be known. In this case, the above formula can be simplified to 
\begin{eqnarray}
\footnotesize
 E(r)  & = &  \frac{V - \delta V}{\ln{\frac{b}{a}}} \cdot \frac{1}{r}\\
&  & (  a \le r \ll b)  \mbox{\rm\ \ \  and} \nonumber \\ 
\delta V \, & = & \,   \frac{\rho e b^2}{4 \varepsilon_0} 
\, =\,  n_p \, g \ t_+ \, \frac{e}{4 \pi \varepsilon_0} \,
\frac{R}{L} \ \ \ .\label{eq:dV}  
\end{eqnarray}
\normalsize
The electric field in the avalanche region behaves
as if the effective voltage at the wire was $ V - \delta V$.\\
Taking now a model for the gas amplification where the gas gain is a
function of the applied voltage, one should be able to describe the
dependence of the gas gain reduction as a function of the rate.
We will use the Diethorn formula \cite{diethorn} 
\footnotesize
\begin{equation} 
g = \exp{\left[ \frac{V \ln{2}}{\Delta V \ln{\frac{b}{a}}} \cdot
\ln{\frac{V}{K \frac{p}{p_0} a \ln{\frac{b}{a}}} }\right] } \ \ .
\label{eq:gain} \end{equation}
\normalsize
\begin{figure}[t]
\centerline{\psfig{figure=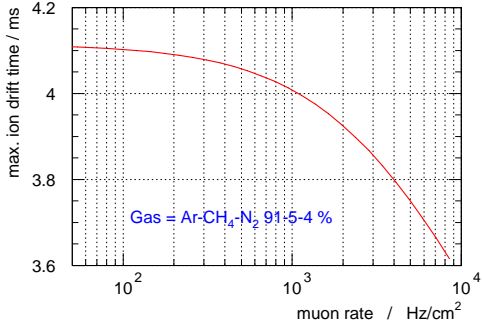,width=\columnwidth}} 
\caption{\protect\footnotesize Change of the maximum ion drift time
with rate.  Calculation using (\ref{eq:newt+}). }
\label{fig=iondrift}
\end{figure}
The Diethorn model has two para\-meters $\Delta  V$ and $K$ which are
characteristic for every gas mixture. $\Delta V$ is the
potential difference an electron passes between two successive
ionisations and $K$ is the critical value of $ E/p $  where ionisation
starts. $ p/p_0$ is the ratio of the actual pressure $ p $ to  the
reference pressure $ p_0$ where K was determined. \\
For calculating the actual gas gain at rate $ R/L $, one now has to
replace $ V $ by $V-\delta V$ in (\ref{eq:gain}) where $\delta V$ is
given by (\ref{eq:dV}). $\delta V$ depends on the gain and
vice versa. The two equations have been solved in an iterative way to get
a self consistent gain for every rate.\\
One more correction has to be made for high rates: it was ignored in (\ref{eq:t+}) that the ion
drift time $t_+$ depends on the rate. For a better approximation of
$t_+$ in (\ref{eq:dV}), one can insert into (\ref{eq:mue})
the electric field (\ref{eq:newfield}) which leads to the formula
\footnotesize
\begin{equation} t_+ = \frac{p \varepsilon_0}{\mu \rho e} \cdot \ln {
\left[ 1+ \frac{\rho e  b^2 
\ln{\frac{b}{a}}}{2 \varepsilon_0 \left( V - \delta V\right)} \right] } 
\label{eq:newt+} \  . \end{equation} 
\normalsize In the limit of rate $ \rightarrow
0$, this is identical to (\ref{eq:t+}). The change of $t_+$ with rate
is plotted  in fig.\ref{fig=iondrift} for the gas Ar-CH$_4$-N$_2$ 91-5-4 \%.\\

\section{Experimental Setup}
\begin{figure}[b]
\centerline{\psfig{figure=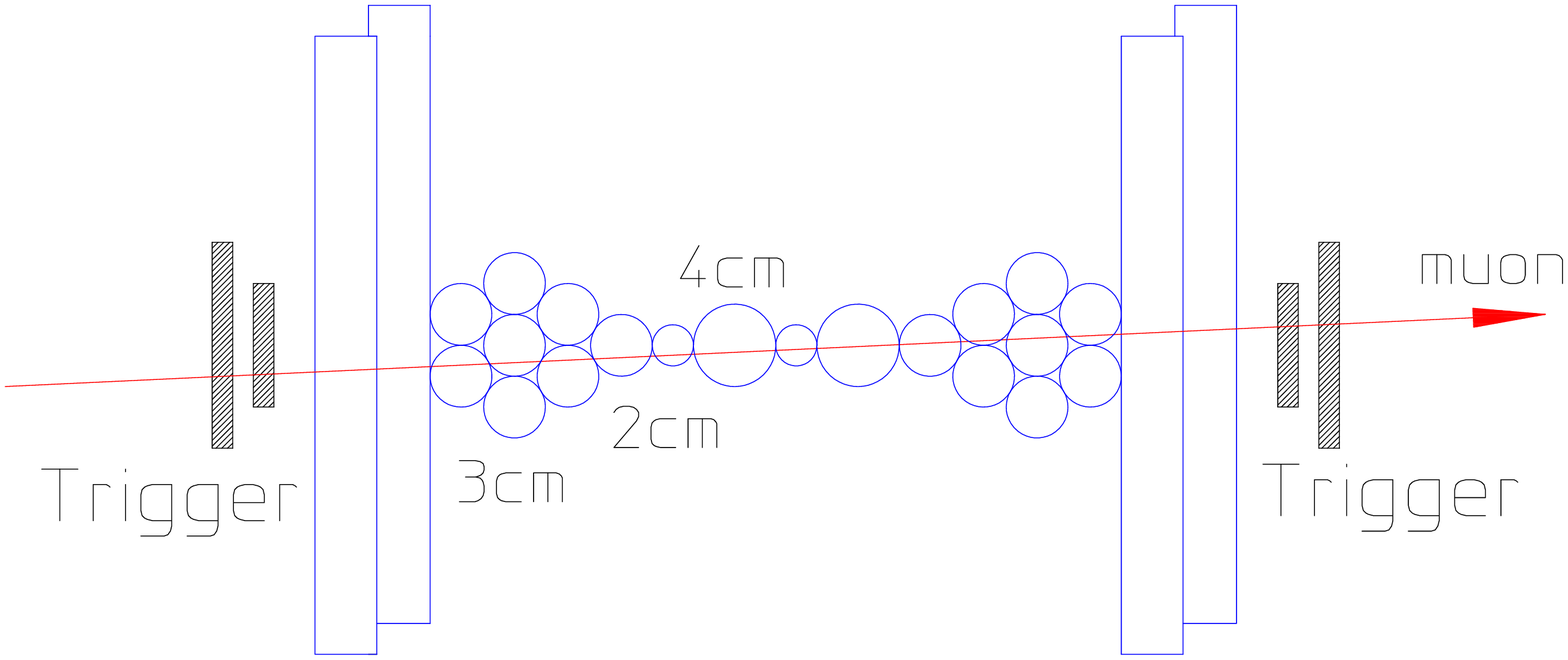,width=\columnwidth}}
\vspace{2mm}
\centerline{\psfig{figure=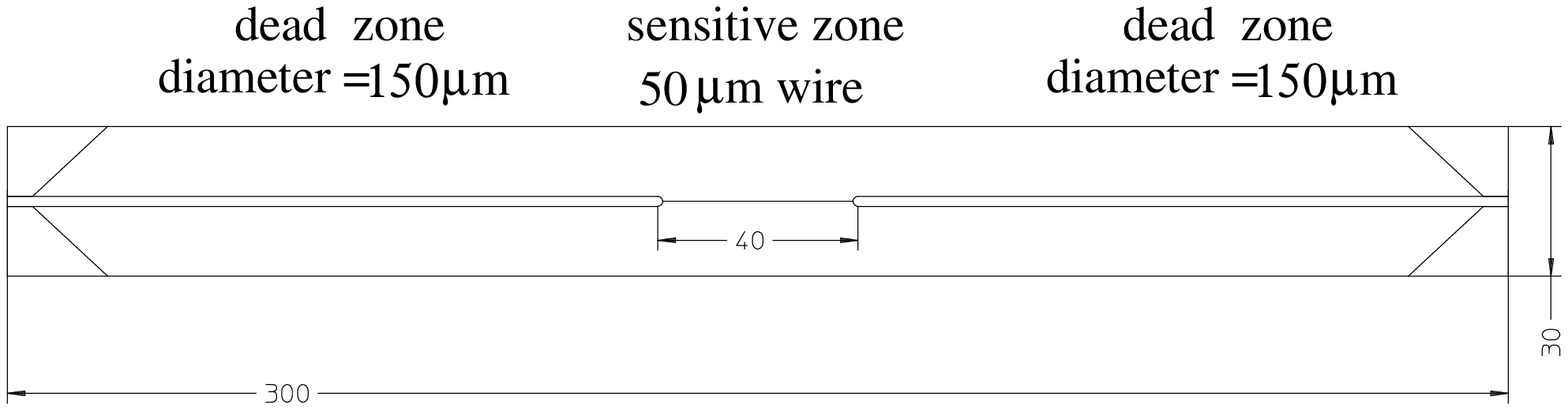,width=\columnwidth}}
\caption{ \protect\footnotesize Top: schematic drawing of the chamber, view
 from top. Bottom: setup of a single tube. The wire is
silver coated to avoid gas amplification except for 4\,cm in the
middle part. }
\label{fig=schematic}
\end{figure}
\begin{figure*}[t]
\centerline{\psfig{figure=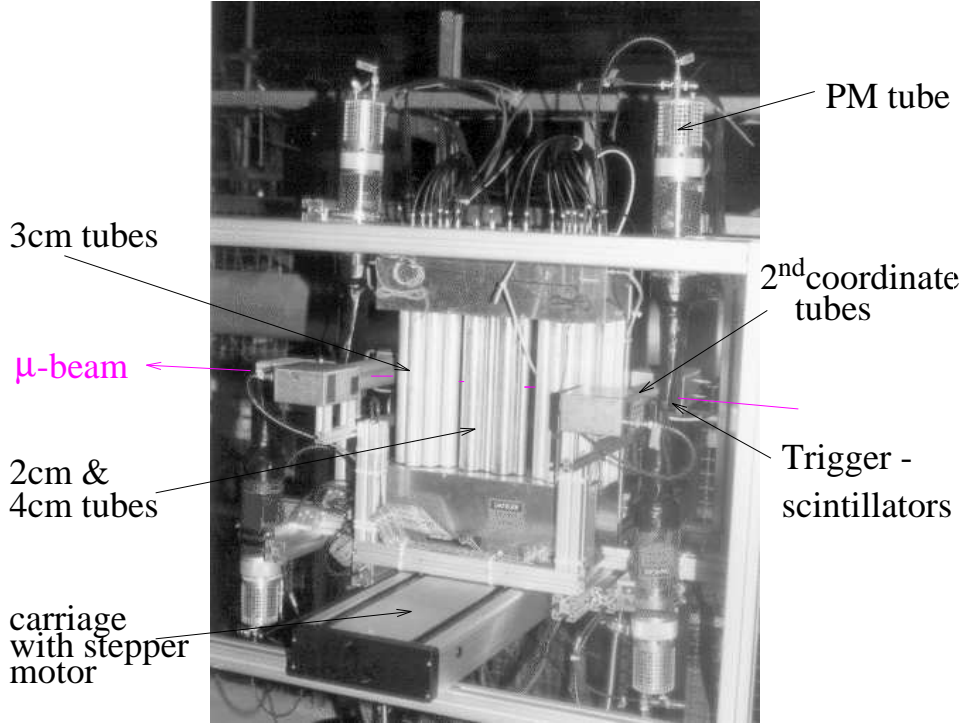,width=\textwidth}}
\caption{\protect\footnotesize Photo of the apparatus in the 190\,GeV
muon beam at CERN.  One can see two bundles of 8\,tubes with 3\,cm
diameter, two tubes with 2\,cm and two tubes with 4\,cm diameter. In
order  to measure the second coordinate, there are two tubes
perpendicular to the others at each side of the chamber. The trigger
is very small, covering only an area of 1\ $\times$\ 6\,cm$^2$ in the
sensitive zone of the tubes.} 
\label{fig=photo}
\end{figure*}
To measure effects of high rates, a specially designed chamber was
built for operation in the M2 muon beam at CERN.  The task of this
setup was to see only effects coming from space charges and to avoid
measuring electronic effects such as saturation of preamplifiers,
baseline shifts  etc.
A schematic drawing and a photo are shown in figs. \ref{fig=schematic}
and \ref{fig=photo}, respectively. The
chamber consists of 24 drift tubes: two bundles of eight tubes
with 3 cm diameter, two tubes with 2 cm and two tubes with 4 cm
diameter.  The tubes are glued together precisely, and the two bundles
are staggered to improve  track fitting.\\
For measuring the second
coordinate, four tubes are placed perpendicular to the others.  All
tubes are 30\,cm long, but the active length is only 4\,cm in the middle
part of the tubes for the precision coordinate and 10\,cm for the
second coordinate (fig. \ref{fig=schematic} bottom). In order 
to passivate the rest of the tube, the wires have been coated
galvanically with silver to increase the wire diameter from 50\,$\mu$m
to 150\,$\mu$m in order to avoid gas amplification. The trigger is made of a
coincidence of four scintillation counters which cover an area of 1\,cm
(along the wires) $\times$ 6\,cm (= two tubes in height) and are placed
in the middle of the active part of the tubes.\\
The whole setup can be moved horizontally by a stepper motor within a
range of 1\,m from the centre of the muon beam. By moving the
apparatus with respect to the beam one can adjust the trigger
rate from 10\,Hz/cm$^2$ up to more than 10\,kHz/cm$^2$. 

\section{Data Readout and Analysis}
Whereas in  {\sc Atlas} the tubes will be read out with TDCs, here the
full analog pulse information was read out with 250 MHz flash ADCs
\cite{fadc}. This is necessary for measuring pulse height reductions
and to be independent of a fixed discriminator threshold.\\ 
\begin{figure}[t]
\centerline{\psfig{figure=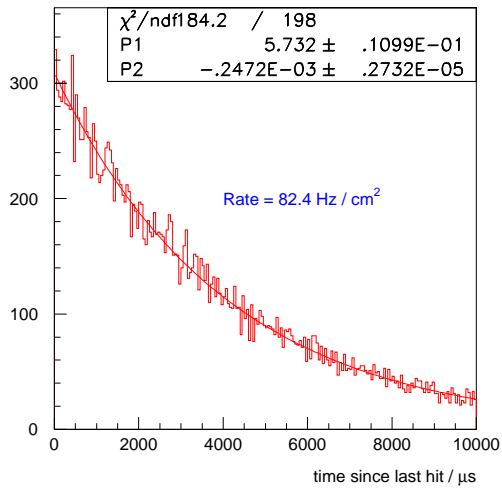,width=\columnwidth}}
\caption{\protect\footnotesize Time differences between two hits in a
tube. Fitted  to the data is an exponential function $ f(t) =
\exp{\left(p_1  + p_2 t \right)} $. }
\label{fig=zeitvert}
\end{figure}
An important property of the data acquisition was the ability to
record  at least 8
successive tracks without dead time. This is necessary for the 
calculation of time distances between consecutive hits 
from which the particle rate is calculated. It is also used to observe
effects coming from event-to-event fluctuations where the "history" of
the tube has to be known. Thus it can be studied how previous hits
influence the properties of the detector for the following hits. \\ 
In order to achieve this readout without dead time, a trigger logic
has been set up that does not need any readout cycle
between the first and eighth event trigger. The memory depth of the
flash ADC is 2048 bytes per channel corresponding to 8192 ns in time
which is large enough to store the waveforms of  8 pulses.
The digitization of the flash ADCs is  started for each new muon
trigger and  stopped after 1\,$\mu$s.\\
\begin{figure*}[t]
\centerline{\psfig{figure=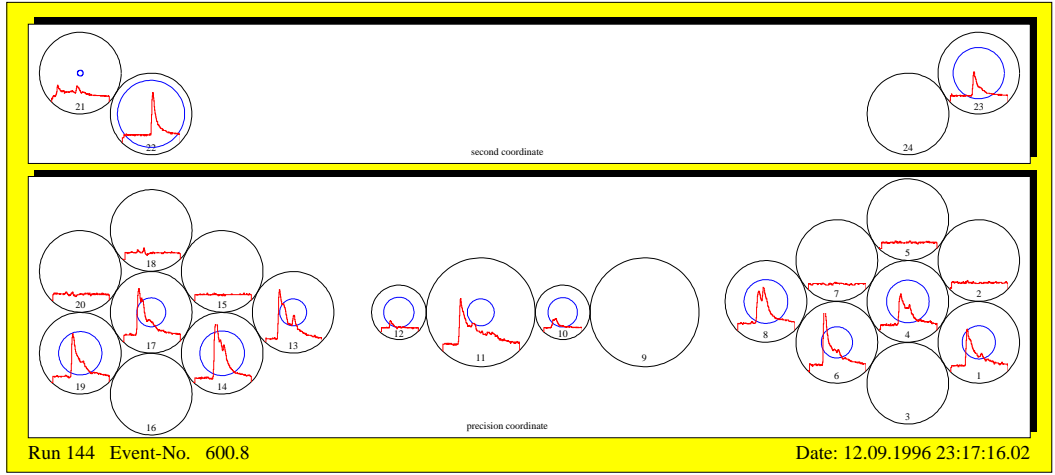,width=1.04\textwidth}}
\caption{\protect\footnotesize Event display of a muon traversing the
chamber. The upper part shows the second coordinate tubes. A circle
inside a tube indicates the reconstructed drift distance if a pulse was
found. The empty tubes have not been read out.}
\label{fig=eventdisplay}
\end{figure*}
The time differences between consecutive triggers were measured with a
setup of slow TDCs (\mbox{5 MHz} clock) covering a time 
range of $\approx$ 20 ms.
With this information it is possible to com\-pute for every tube the
time differences to the last hit. The
distribution of these times is exponential (fig. \ref{fig=zeitvert}).
The mean value is   $ \langle t \rangle = \frac{-1}{p_2}$, in
agreement with the number of triggers per 
second counted with a scaler.  With this measurements we know both,
the time difference between two consecutive hits and (from the  second
coordinate measurements) the spatial distance of the muon tracks
along the wire. \\
A typical size for the data that had to be read out after every cycle (8
events) was 40\,kB. The 
data were processed by a VME processor operated under OS\,9 and stored on
an exabyte tape. To maximize the readout speed, all time critical
routines have been written in assembler code.  The maximum readout
rate that could be achieved (if the particle rate was high enough) was
$\approx 800$ events within one spill of the CERN SPS accelerator
(2.4\,s beam followed by a  
pause of \mbox{12 s}) and was limited by the speed of the tape drive.
The time 
needed to take a high statistics run (800.000 triggers) at highest rates
was 4 hours.\\
A slow control system was installed to monitor the relevant parameters
such as temperature, gas flow, gas mixture, pressure and the high
voltages of the tubes.  \\
The measurement of drift times with this kind of flash ADC needs a special
setup.
The time difference between incoming start signal and start of the
digitization can vary randomly between 0 and 12\,ns. In order to get a
precise time 
marker, the trigger signal starting the readout was delayed by 800
ns and fed into the test input of the preamplifiers FBPANIK-04
\cite{panik}. Thus, every drift chamber pulse
was followed by a reference pulse coming always with the same delay
with respect to
the muon trigger. The time difference between
the pulse and the reference pulse is the drift time plus a constant
offset t$_0$.\\
The time bins of the flash ADCs have a width of 4 ns. To
get a better time resolution of the pulse, the DOS (difference of
samples)  method  was used which was developed
for the analysis of the flash ADC data of the OPAL
experiment\cite{dos1,dos2}.  This
method is numerically rather simple but very efficient in both finding
pulses and calculating the drift time with an accuracy better than
1\,ns. First, the pulse shape is
differentiated by subtracting the pulse heights of successive
bins. Then pulses can be found if the differentiated pulse reaches a
certain threshold. Finally the drift time is calculated as the
weighted mean of the differentiated pulse around
the position of the maximum of the differentiated pulse. \\
An event display of a muon passing the chamber is shown in fig.
\ref{fig=eventdisplay}.\\

\section{Gain Reductions}
\begin{figure}[t]
\centerline{\psfig{figure=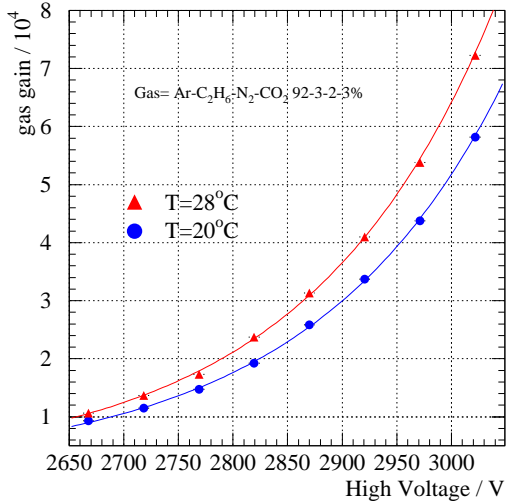,height=\columnwidth}}
\caption{\protect\footnotesize Measurements of the gas gain for
Ar-C$_2$H$_6$-N$_2$-CO$_2$ 92-3-2-3\,\%. For precise measurements of the
gain it is necessary to keep the temperature constant. The solid lines
show the Diethorn fits to the data taken at 20$^{\rm o}$C and 28$^{\rm
o}$C. }   
\label{fig=gainmeas}
\end{figure}

In order to measure the correct gain drop that is only due to particle
rate one has to take care of several effects that influence the
measured pulse height.  First of all, only data from the same tube are
compared to exclude systematic errors like dependence on
preamplifiers or slightly different gas mixtures. \\
One important quantity is the temperature of the gas. The
pulse height increases with temperature if the pressure is kept
constant (fig.\,\ref{fig=gainmeas}). The order of magnitude of this
effect is some percent per Kelvin. 
\begin{figure}
\centerline{\psfig{figure=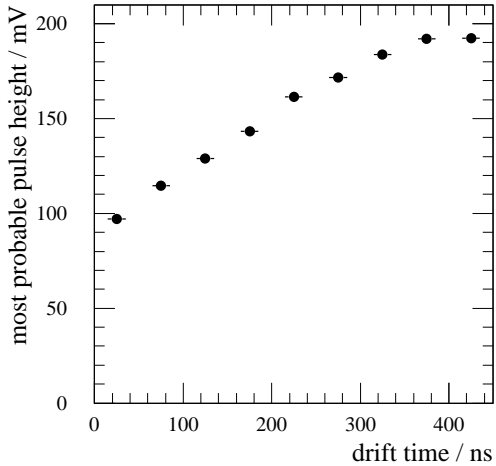,height=0.95\columnwidth}}
\caption{\protect\footnotesize The pulse height depends on the drift
distance. This is due to the different pulse shapes for 
pulses with short and long drift distances. Measurement with
Ar-CH$_4$-N$_2$ 91-5-4 \% at 3 bar and a gas gain of $2
\cdot 10^4$.    }
\label{fig=ph_dt}
\end{figure}
\begin{figure}
\centerline{\psfig{figure=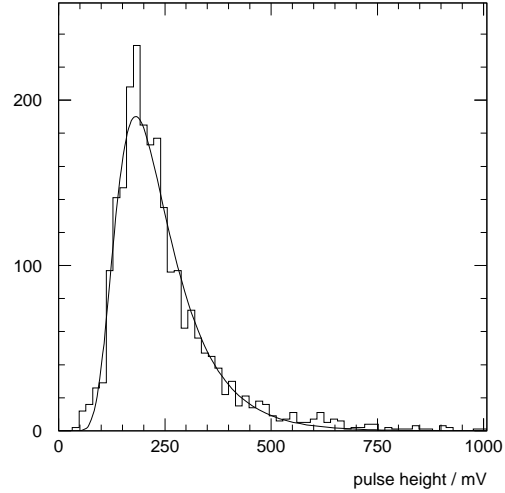,width=\columnwidth}}
\caption{\protect\footnotesize Histogram of the maximum pulse heights
of measured pulses. A Landau function 
is fitted to this  distribution to determine the most probable
pulse height.}
\label{fig=landaufit}
\end{figure}
Since the data were taken in a non-air-conditioned hall with temperature
variations of up to 5\,K within one measurement, temperature
corrections of the pulse height had to be done. Gas gain measurements at
different temperatures have been done with the setup described in
\cite{karen} for the gases that were tested in beam measurements. Thus
it was possible to apply the correct 
temperature corrections and to get the Diethorn parameters needed for
the calculation.\\
However, the most important point is to compare only pulses with equal
drift times for the primary electrons.
As can be seen in 
\mbox{fig. \ref{fig=ph_dt}}, the measured pulse height increases with
drift time (with distance from the wire). Pulses with a short drift
time have a lower maximum and a longer tail compared to those with
long drift time -- although the primary ionisation is lower due to the
shorter track length inside the tube.  Therefore, a cut on the drift
time was applied to ensure that only pulses with approximately the
same drift distance were compared.\\ 

\begin{figure*}[t]
\centerline{\psfig{figure=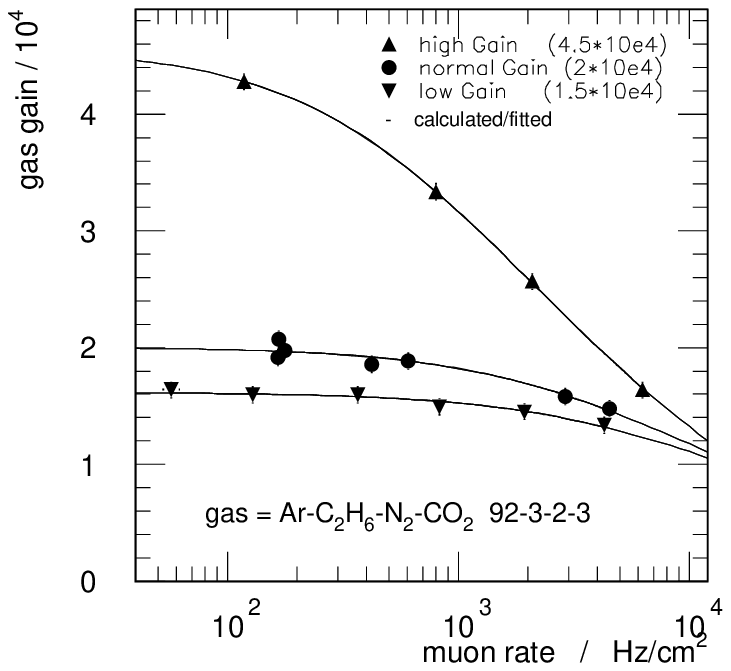,width=0.48\textwidth}\hfill
         \psfig{figure=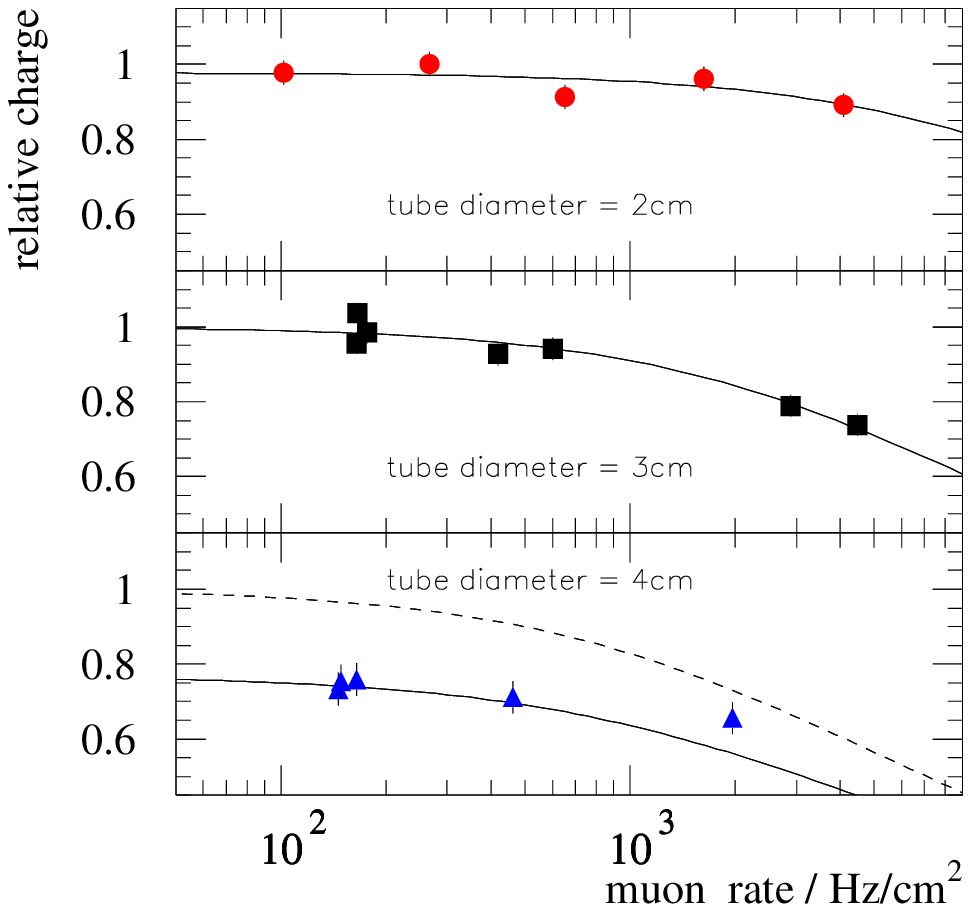,width=0.48\textwidth}} 
\caption{\protect\footnotesize Left: Degradation of the gain with
rate. Right: gain reduction vs rate for 2, 3 and
4\,cm tubes. The relative charge 1 refers to the charge
deposit for a muon track in  a 3\,cm tube operated at a
gas gain of $2\cdot 10^4$. The markers show measured data, the lines
are calculations. The dashed line in the right plot is an
extraplolation from the measured data to rel. charge=1. All
measurements were done with the gas 
Ar-C$_2$H$_6$-N$_2$-CO$_2$ 92-3-2-3\,\%.}   
\label{fig=height_rate}
\label{fig=height_tuberad}
\end{figure*}
After
these cuts, the maximum of each pulse was determined and histogramed.
Because the muons are minimum ionising particles, the distribution of
the pulse heights follows a Landau function \cite{sauli} which was
fitted to the data. This is shown in fig. \ref{fig=landaufit}.
To describe the distribution of the pulse heights, the most probable
pulse height (which is one parameter of the fit) was used and not 
the mean value of the data sample (which is bigger). 
In the following pulse height means the most probable pulse height of
the Landau fit.\\ 
Fig. \ref{fig=height_rate} shows the behaviour of the
pulse height with increasing rate for three different settings of the
high voltage, corresponding to gas gains of $ 1.5 \cdot 10^4 $,  $ 2
\cdot 10^4 $ and $ 4.5 \cdot 10^4 $ at zero rate. \\ 
The solid lines show a calculation of the 
pulse height using  the Diethorn formula (\ref{eq:gain}) to calculate
the gain drop. 
For the calculation of the electric field
(\ref{eq:newfield}) and the actual gain, several iterations of the
calculation of $\delta$V (\ref{eq:dV}), the gas gain (\ref{eq:gain})
and the maximum ion drift time (\ref{eq:newt+}) are necessary to
obtain an accurate value for the charge density $\rho$
(\ref{eq:rho}).  To compare measured pulse height and calculated gain,
one needs to know the ion mobility  which was fitted to $\mu =
1.5\,\frac{cm^2 bar}{V s}$ and found to be consistent with the value of 
1.53\,$ \frac{cm^2 bar}{V s}$  for Ar$^+$ in Ar from
\cite{McD,landolt}.\\
One can see that an  initially high gas gain drops
rather fast with increasing rate, going down to 30\% of its initial
value at 10\,kHz/cm$^2$.  For nominal {\sc Atlas} conditions with a gas gain
of 2$\cdot 10^4$ and photon rates $\le$\,100\,Hz/cm$^2$ (corresponding
to a muon rate of 200\,Hz/cm$^2$) the behaviour is rather
uncritical, the reduction is some 10\%.\\
However, at very high rates,
the pulse height is rather independent of the high voltage, the
electric field is reduced to the same limiting value.
The order of magnitude of the gain drop with rate is approximately
the same for different gases, but the gain drop is larger for gases with
low working point (low high voltage for the same gas gain). In that
case, the ion drift time is longer and so the space
charge density is higher.\\
The same study was done for the 2\,cm and 4\,cm tubes. Because the
primary ionisation of a muon is proportional to its path length inside
the tube, the signals of a 4\,cm tube are higher than those of 3\,cm and
2\,cm tubes if all are operated at the same gas gain. In order to
compare the behaviour of the different tubes, measurements at  the same
charge per track (not at the same gas gain) at zero rate were compared. 
For easier comparison, the charge for a 3\,cm tube at a gas gain of
$2\cdot 10^4$ is set to 1 in fig.\,\ref{fig=height_tuberad}. For the
4\,cm tube, no measurement was available for charge=1, so a
calculation is shown (dashed line) together with a measurement for
lower charge. One can clearly see that the
gain drop is increasing for bigger tubes at the same irradiation. It is
almost negligible for the 2\,cm tube, but even for the 4\,cm one, the
gain drop is in the order of 20\,\% at a muon rate of 1\,kHz/cm$^2$
and the nominal gas gain.\\
The gain reduction could be totally compensated by increasing the high
voltage by $\delta$V (\ref{eq:dV}) corresponding to the actual rate
-- assuming that the irradiation is homogeneous along the tube.

\section{Changes in the Electron Drift Time}
\begin{figure}[t]
\centerline{\psfig{figure=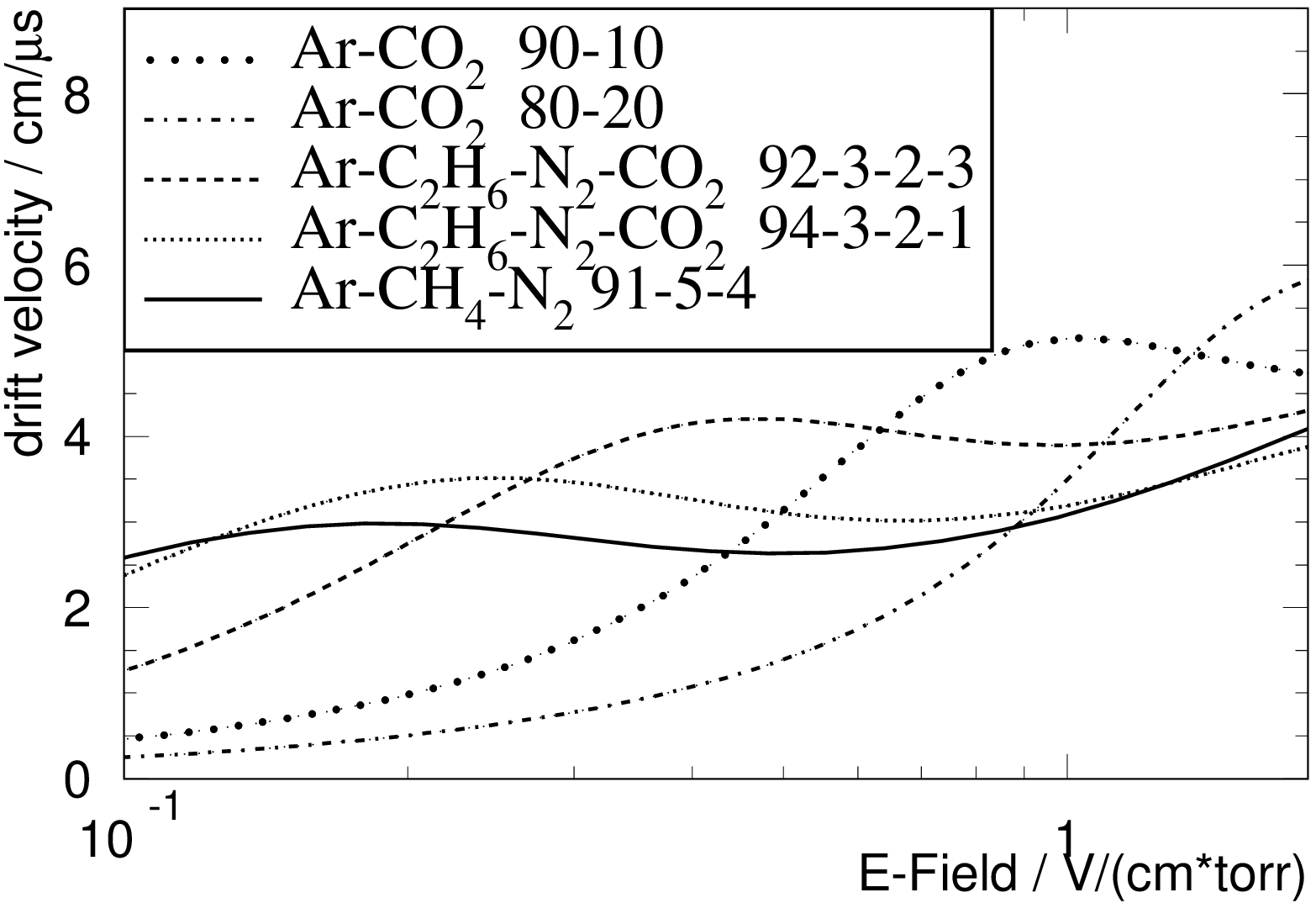,width=\columnwidth}}
\centerline{\psfig{figure=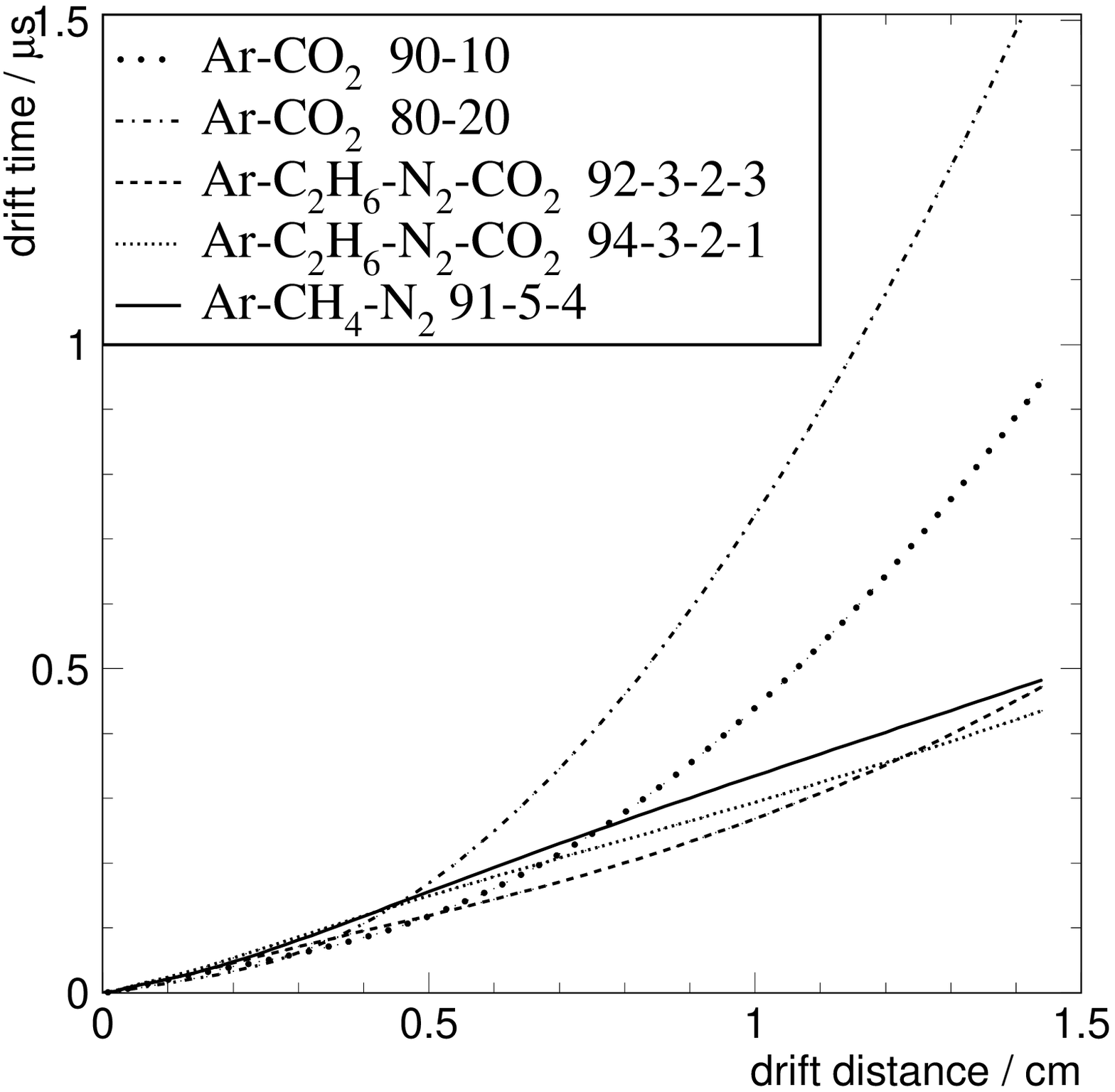,width=\columnwidth}}
\caption{\protect\footnotesize Top: comparison of the drift velocity for
different gases. The typical electric field inside a drift tube
ranges from 0.15 V/(cm torr) at the tube wall of a 3\,cm tube  up to
1.8 V/(cm torr) at 1\,mm from the wire. Bottom: r-t
relations for these gases. 
Gases with linear r-t relation have a
rather constant drift velocity in the relevant region of E/p compared
to the other gases. }
\label{fig=velo}
\end{figure}
\begin{figure}[bt]
\centerline{\psfig{figure=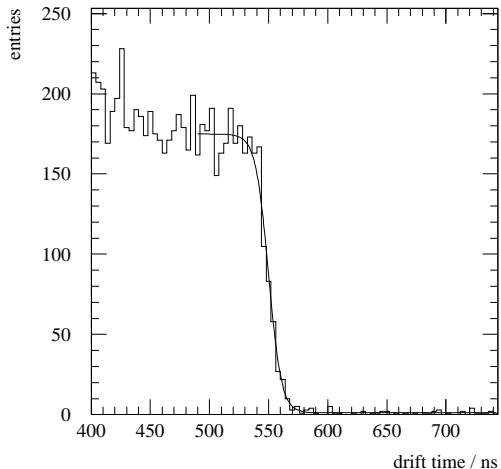,width=\columnwidth}}
\caption{\protect\footnotesize Fit of the maximum electron drift time.
 A Fermi function  is fitted to the end of the
drift time distribution.    }
\label{fig=fermifit} 
\end{figure}
\begin{figure*}[t]
\centerline{\psfig{figure=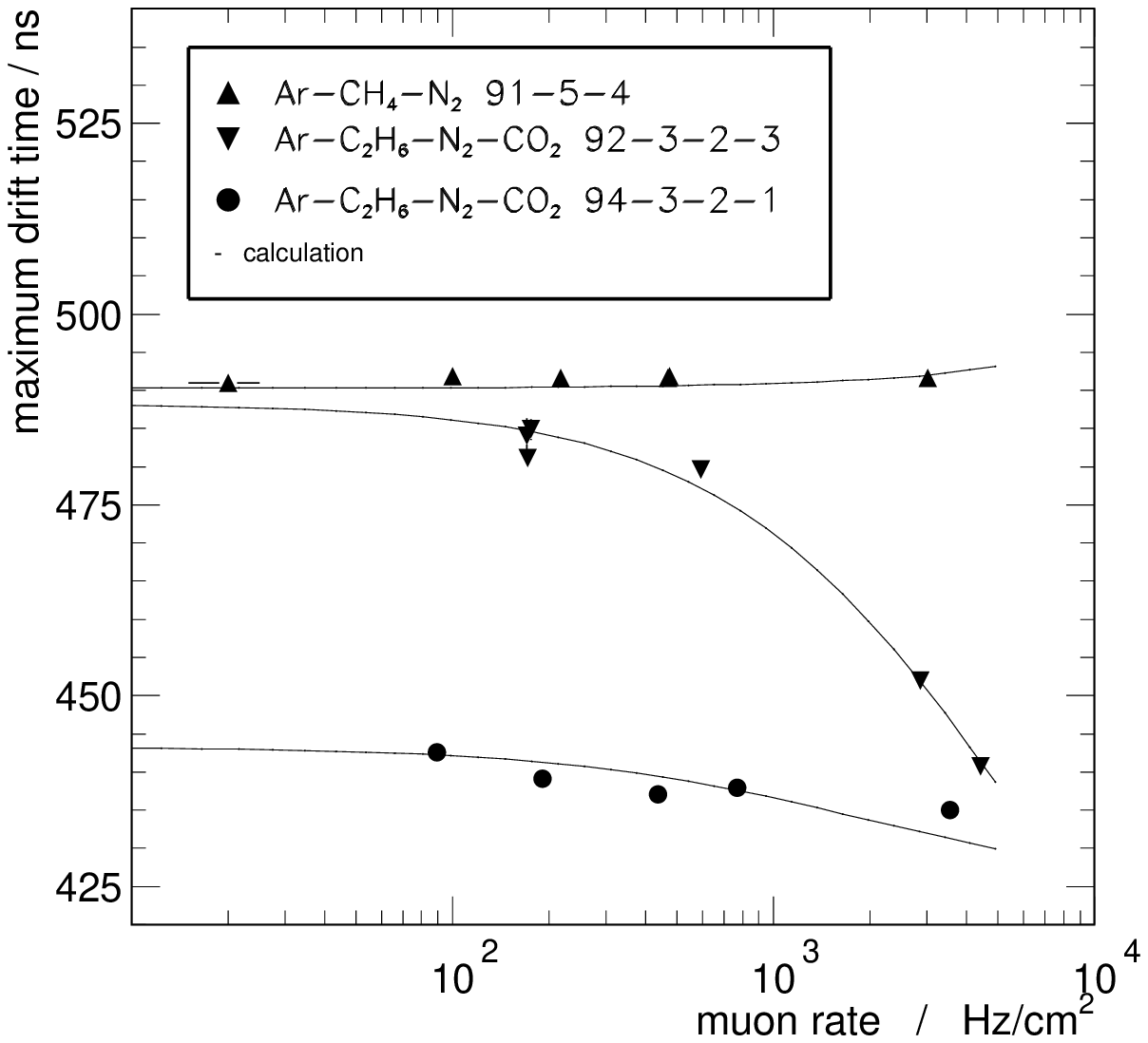,width=0.49\textwidth}
\hfill \psfig{figure=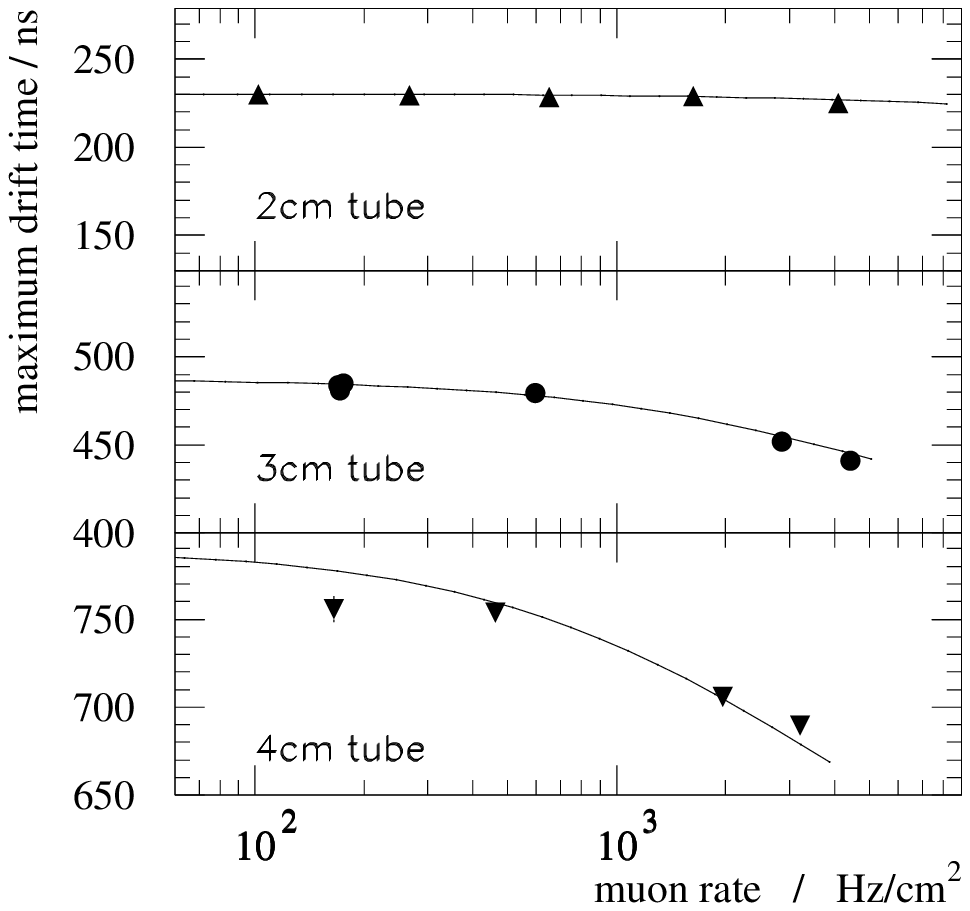,width=0.49\textwidth}}
\caption{\protect\footnotesize Left: Comparison of measured change in
the maximum electron drift time and calculation for different gases
(3\,cm tubes) 
Right: Comparison of the change of the maximum
drift time versus rate for tubes of 2, 3 and 4\,cm diameter.  All
measurements in the right plot were done with the gas Ar-C$_2$H$_6$-N$_2$-CO$_2$ 92-3-2-3\,\%.}
\label{fig=drift_calc}
\end{figure*}

The electric field for drifting electrons is changed with rate
according to (\ref{eq:newfield}). Whereas the field changes are more
or less the same for different gases at the same rate (at least if
they have  
to be operated at the same high voltage), the effects to the electron
drift times are different. The drift time is given by
\begin{equation} t(r) = \int_a^r{\frac{d r'}{v(E(r'))}} \ \ \ . 
\label{eq:t_r}
\end{equation}
The change in drift time depends on the slope
of the drift velocity $v(E)$ with the electric field. 
A gas becomes faster with rate if
the slope is positive and slower if it is negative.\\ 
One can see here
that in order to keep the drift time changes small, one has to look for
gases with constant drift velocity, so called linear gases (linear
because in this case, the r-t relation is a straight
line). Fig. \ref{fig=velo} shows the drift velocities and the
corresponding r-t relations for the three gases that have been studied
in this experiment: the linear gases Ar-CH$_4$-N$_2$
91-5-4\,\%, and Ar-C$_2$H$_6$-N$_2$-CO$_2$ 94-3-2-1\,\% and
the rather non-linear  Ar-C$_2$H$_6$-N$_2$-CO$_2$ 92-3-2-3\,\%. For
comparison, 
the very non-linear gases Ar-CO$_2$ 80-20\,\% and Ar-CO$_2$
\mbox{90-10\,\%}  are added.\\
To study the effects of the modified  electric field, we look
at the change of the maximum electron drift times at different
irradiation rates.\\ 
The maximum drift time can be calculated using 
(\ref{eq:t_r}) with $r =$ radius of the tube. 
To do this, one needs to know the drift velocity $v(r)$ of the gas
mixture. The {\sc Magboltz} \cite{biag89} program
was used to 
calculate the drift velocity as a function of the reduced electric field
$v(\frac{E}{p})$.
Eq. (\ref{eq:newfield}) was taken to
calculate the correct electric field $E(r)$ and then the new r-t
relation was integrated using (\ref{eq:t_r}).\\
 For using the equation
(\ref{eq:newfield}) the same numbers for the charge density (and the
ion mobility) were used as for the calculation of the gain drop in the
previous section. 
The calculated maximum drift time is shown as lines in 
fig. \ref{fig=drift_calc}, together with the measured points.\\
The maximum electron drift time of the measured data can be determined
from the falling edge of the distribution of the drift times. Because of
diffusion and vanishing ionisation at the tube wall, the tail is
smeared. To get a precise and reproducible number, a Fermi function 
$ f(x)=\frac{p_1}{exp((x-p_2)/p_3)+1} + p_4 $
has been fitted to the distribution (\cite{italofit}). The four
parameters of the fit 
are the upper level ($p_1$), the time at half height ($p_2$), the slope of
the fall ($p_3$), and an offset ($p_4$). Parameter $p_2$ is taken as
the maximum drift time. 
The physical maximum drift
time may be different from this value by a constant offset. However,
we are primarily interested in changes of the maximum drift time.
An offset between measurement and calculation was fitted and
compensated.\\
Three gases are shown in the left plot of fig.\,\ref{fig=drift_calc}
 for a 3\,cm tube: the linear
gas Ar-CH$_4$-N$_2$ 91-5-4\,\%, the non-linear gas with the same
maximum drift time Ar-C$_2$H$_6$-N$_2$-CO$_2$ 92-3-2-3\,\% and the
linear gas Ar-C$_2$H$_6$-N$_2$-CO$_2$ 94-3-2-1\,\%. For the first one,
the maximum drift time is stable up to 
very high rates with a tendency to increase with rate. The non-linear
one shows already differences of several nanoseconds at rather low rates
(\mbox{100 Hz/cm$^2$}). 
For comparing the 2\,cm and the 4\,cm tubes to our standard 3\,cm one,
fig.\,\ref{fig=drift_calc}(right) shows the change of the maximum
 drift time with rate in each case.\\
The data points for the 3\,cm tube are the same as in
the left plot for the gas Ar-C$_2$H$_6$-N$_2$-CO$_2$
92-3-2-3\,\%. The solid lines show the calculation done in the same
way as for the left plot. The datasets used to compute the right plot
 are the same as those for the right plot of fig.\ref{fig=height_tuberad}.\\
As for the gain reduction, one can see here that space charges become
more and more important if the tube diameter is increased. The 2\,cm
tube might be operated with a non-linear gas if that was already
considered impracticable for a 3\,cm tube. For the 4\,cm tube, the drift
characteristics are dramatically changed at particle rates of
1\,kHz/cm$^2$ or more. It is surely not advisable to use it for high
precision measurements at high rates.\\
Surprisingly, both measurement and calculation show that the gas
with the most linear r-t relation is not the one that is most
insensitive to rate. 
The reason for this is the fact that the operation point of 
Ar-C$_2$H$_6$-N$_2$-CO$_2$ 94-3-2-1\,\% is
much lower compared to the others (2575\,V instead of 3300\,V for 
Ar-CH$_4$-N$_2$ 91-5-4\,\% and a 3\,cm tube. This lower voltage has
two effects:  
\begin{itemize}
\item the amount of space charge is proportional to the inverse of the
voltage.
$ \rho \propto t_+ \propto \frac{1}{V}$ because of (\ref{eq:rho}) and
(\ref{eq:t+}). 
\item the relative change of the electric field is proportional to the
inverse square of the voltage\\
\begin{figure}[t]
\centerline{\psfig{figure=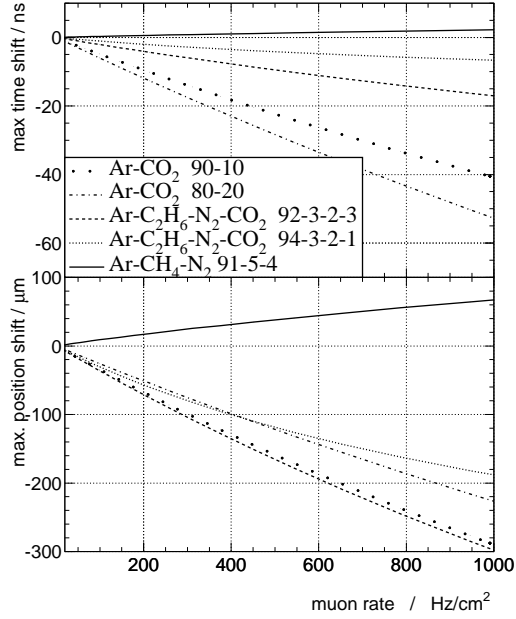,width=\columnwidth}}
\caption{\protect\footnotesize Top: Change of the maximum drift time
with rate. Same calculation as in the previous plot. For comparison,
the very non-linear gases Ar-CO$_2$ \mbox{80-20 \%} and Ar-CO$_2$
\mbox{90-10 \%} are added. Bottom: corresponding shift in the spatial
position.}   
\label{fig=terr}
\end{figure}
\begin{figure}
\centerline{\psfig{figure=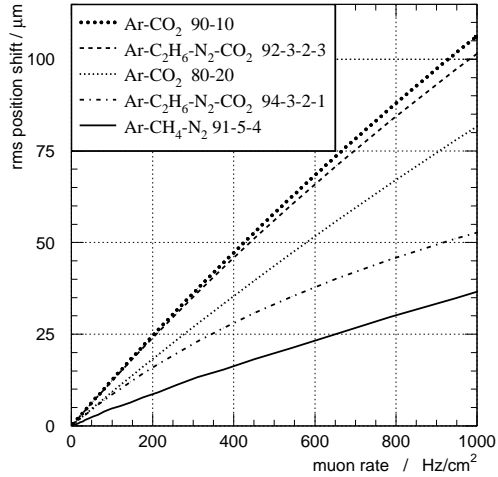,width=\columnwidth}}
\caption{\protect\footnotesize rms of the  position shift for all
radii at given rate. }
\label{fig=rmserr}
\end{figure}
$\frac{\Delta E}{E} \propto \frac{1}{V^2}$ because of
(\ref{eq:newfield}). 
\end{itemize}
Both points show that an optimum gas should have a high operational
voltage. The reason for the rather low working point of this gas is
the use of C$_2$H$_6$ as quencher. By replacing the Ethane with
Methane, the voltage for the same gas gain will increase by 500\,V
keeping the r-t relation almost unchanged. \\
Fig. \ref{fig=terr} shows the same calculation as fig.
\ref{fig=drift_calc} for low rates in a linear scale. Additionally,
calculations for the gases Ar-CO$_2$ 80-20 \% and 90-10 \% are shown
for comparison with non-linear gases. \\
Instead of comparing time shifts it is more adequate to compare the
resulting shift in position. Fig. \ref{fig=terr} bottom shows this
position shift (=the difference between real an assumed position of
the particle) when using the r-t relation suited for rate=0
at the actual rate. This value is plotted for the maximum drift radius
(1.5\,cm) where for most gases this shift has a maximum. 
It is much smaller for smaller drift radii. 
The reason for this is
the slope of the drift velocity with the electric field (fig.\,\ref{fig=velo}). If
it is zero, there is no shift because the drift
velocity stays 
constant. If the slope is monotonous the shift increases with drift
radius. This is the case for most gases.
For the gas Ar-CH$_4$-N$_2$ 91-5-4\,\%, the slope is very low
and the sign changes twice in the relevant range of E/p so the
integrated effect is almost zero for the maximum drift distance.\\
Calculating the rms of the values for all
radii at given rate leads to fig.\,\ref{fig=rmserr}. This is a good
measure for comparing the rate behaviour of different gases.
Typical values are in the order of 50-100\,$\mu$m for a muon rate of
1\,kHz/cm$^2$,  but for linear
gases 30\,$\mu$m can be reached which is almost negligible compared
to the intrinsic resolution of 70\,$\mu$m of the tube.\\
\begin{figure}[tb]
\centerline{\psfig{figure=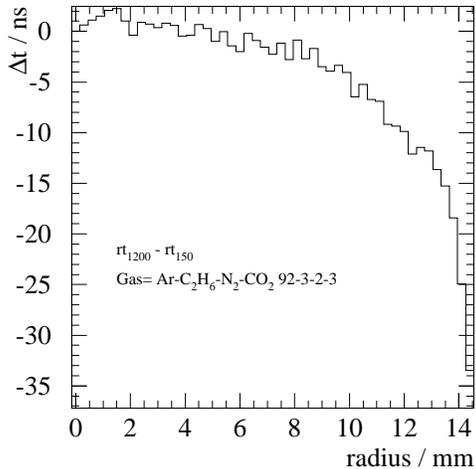,width=\columnwidth}}
\caption{\protect\footnotesize Difference of the rt relations
calculated from data for rates of 150\,Hz/cm$^2$ and
1200\,Hz/cm$^2$.}
\label{fig=rtdiff}
\end{figure}
The position shift does not necessarily decrease the resolution of the
drift tube. If the rate is stable, one can calculate an
adapted r-t relation where the position shift is compensated.
Such a  r-t relation can be calculated directly from the data with
the autocalibration method \cite{autocal}. Fig. \ref{fig=rtdiff} shows the
difference between two r-t relations calculated from data at
rate=150\,Hz/cm$^2$  and rate=1200\,Hz/cm$^2$.

\section{Fluctuations }
\begin{figure}[tb]
\centerline{\psfig{figure=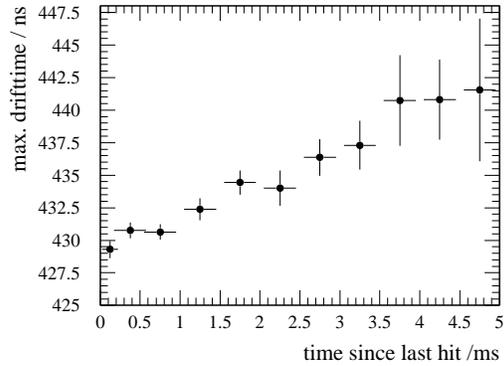,width=\columnwidth}}
\caption{\protect\footnotesize Maximum drift time vs time since last
hit in the tube (gas=Ar-C$_2$H$_6$-N$_2$-CO$_2$ 92-3-2-3\,\%).}
\label{fig=dt_tl} 
\end{figure}

\begin{figure}[tb]
\centerline{\psfig{figure=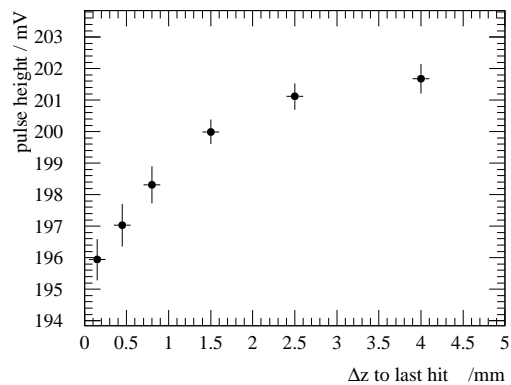,width=\columnwidth}}
\caption{\protect\footnotesize Pulse height reduction versus distance
along the wire to the previous track (gas=Ar-C$_2$H$_6$-N$_2$-CO$_2$ 92-3-2-3\,\%).}
\label{fig=ph_dz} 
\end{figure}

In this section we will show that the resolution
becomes worse at high particle rates even if one uses the
r-t relation  with the position shift compensated. 
It will be shown that the reason for the resolution loss is the
fluctuation of the times between two hits which was ignored up to
now where only the mean value was taken into
account. 
Fig.\,\ref{fig=zeitvert} shows that the  
times between two hits are distributed exponentially. The time
constant of the exponential function is defined by the rate. That
example for rate $\approx $ 80\,Hz/cm$^2$ also shows that for roughly
half of the 
events the last hit was more than 4\,ms ago, which is the time of the
ion drift for a 3\,cm tube. In that case, all space charges are gone
and the tube
behaves as if the rate was zero. On the other hand, we know that for
a rate of 80\,Hz/cm$^2$, the mean position shift is not negligible for
most gases. If the time distance between two events is much shorter
than 4\,ms, one has to take into account all hits within this time
window in order to get the correct r-t relation. This 
is impossible for most applications.\\
With our setup, it is possible to see the influence of previous
hits. Fig.\,\ref{fig=dt_tl} shows how the  maximum
drift time depends on the time since the last event passed for  the
gas Ar-C$_2$H$_6$-N$_2$-CO$_2$ 92-3-2-3\,\%.
All tracks were used that were within one cm along the wire
coordinate. A tighter cut did not change the behaviour significantly.\\
The second example (fig.\,\ref{fig=ph_dz}) shows the pulse height versus
the distance $\Delta$z along the wire to the previous track. One can
clearly see a decrease of the pulse height in the immediate vicinity
($\le$ 1\,mm) of the 
previous track, but the effect is very small ($<$2\%).\\
For a better understanding of the space charge fluctuations, a
simulation study was done and afterwards compared to the measured data.
In a first step, it was investigated, how an ion cloud produced at the
wire evolves with time in three dimensions. A typical charge cloud
consists of 10$^7$ ions (primary ion pairs multiplied by gas gain).
Ions were generated near the wire with an initial radial
coordinate  distributed exponentially within 50\,$\mu$m from the
wire. The position of the coordinate along the wire (z) was
distributed like a gaussian with
$\sigma_z=0.2$\,mm. The same was true for the azimuthal angular distribution
with $\sigma_\phi=60^{\rm o}$. The 
ions were represented by 1000 points, each one carrying 1/1000 of the
total charge. The effective electric field (coming from the applied
voltage and the field of the point charges) was calculated at each of
these points. Now, one can calculate the drift velocity of the ions and
from this the new position after the time dt.  
To describe the status of the whole cloud at the time T 
after creation, the mean value and the rms of the coordinates r, $\phi$
and z were calculated.\\
In the simulation, the extension of an ion cloud does not exceed 
$\sigma_z=0.5$\,mm) over the whole drift path.
But within a z range of 5\,mm, field distortions cause changes in the
drift time of more than 1\,ns for non-linear gases
(e.g. Ar-C$_2$H$_6$-N$_2$-CO$_2$ 92-3-2-3\,\%). 
On the 
contrary, for the pulse height (or the gain) only the electric field in
the immediate vicinity of the wire is relevant. The field distortions
coming from an ion cloud that is some mm away are negligible compared
to the very high field near the wire.\\
\begin{figure}[tb]
\centerline{\psfig{figure=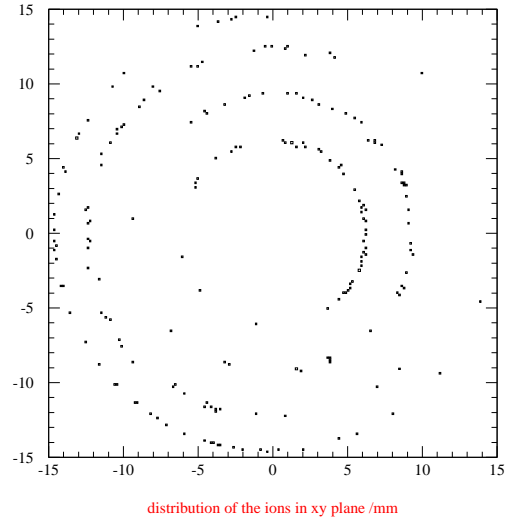,width=\columnwidth}}
\caption{\protect\footnotesize Sample of the distribution of positive
ions in the tube. One can see five rings resulting from previous hits. }
\label{fig=ioncloud} 
\end{figure}
In a second step of the simulation, the time distances T$_i$ between 
two hits were
generated according to the given rate. Ions were generated at
positions according to the 
results of step one for all previous tracks within the time of the ion
drift. Fig \ref{fig=ioncloud} gives an example of the distribution of
space charges coming from different events. 
Now, electrons were put at radii r=0.5\,mm ... r=14.5\,mm (or
the respective tube radius for the 2 and 4\,cm tubes)
and the drift in the (static) field resulting of all relevant space
charges and the applied voltage was calculated. The time needed to
arrive at the wire was histogramed for each radius.\\
\begin{figure}[tb]
\centerline{\psfig{figure=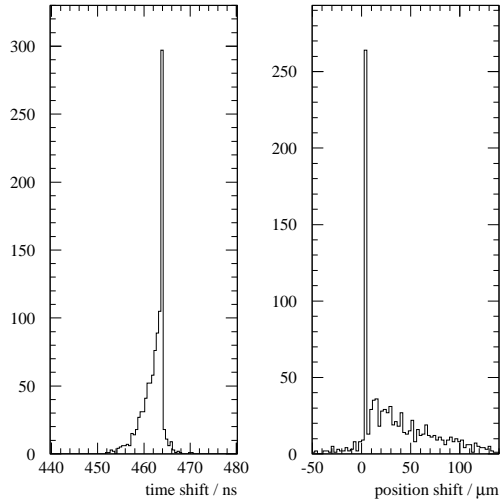,width=\columnwidth}}
\caption{\protect\footnotesize Left: distribution of drift times for
radius=14.5\,mm for a particle rate of 200\,Hz/cm$^2$. Right: position
shift between real position and the one calculated with the r-t
relation. Simulation done for the gas Ar-C$_2$H$_6$-N$_2$-CO$_2$
92-3-2-3\,\%} 
\label{fig=shiftdemo} 
\end{figure}
Fig. \ref{fig=shiftdemo} (left) shows the distribution of these times
for radius=\,14.5\,mm. The right plot shows the position shift
resulting from 
inserting the drift times in the r-t relation (for rate=0) and
subtracting the real position where the electron started. The big,
narrow peak belongs to tracks where no charges were inside the
tube because the last hit was longer than the ion drift time ago. The
distributions become more and more gaussian for higher 
rates.\\
From the histogram of the position shifts, two parameters are
extracted: the mean value which tells how wrong the r-t relation is
for the actual rate and the rms which gives the contribution to 
resolution decrease. This number has to be added quadratically to the
resolution of the drift tube at zero rate to get the correct value
for the resolution at the investigated rate.\\
\begin{figure}[t]
\centerline{\psfig{figure=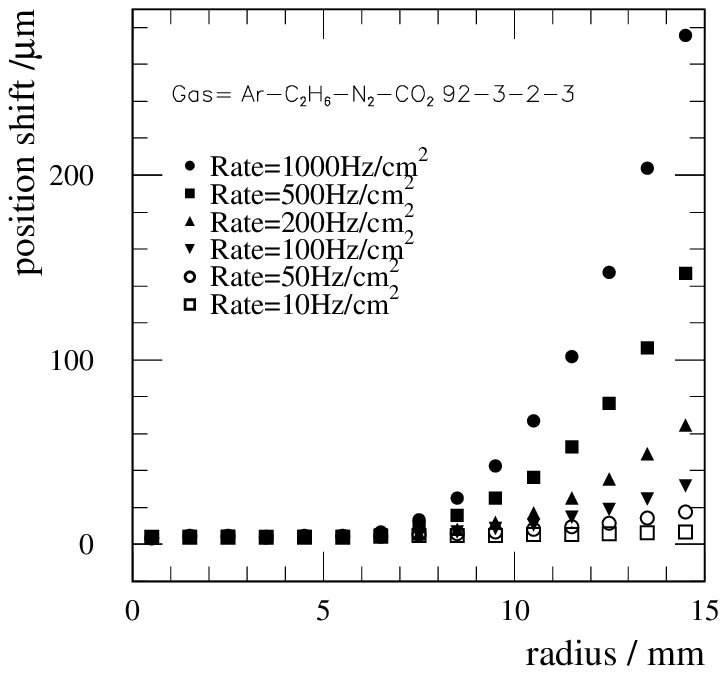,width=\columnwidth}}
\vspace{2.mm}
\centerline{\psfig{figure=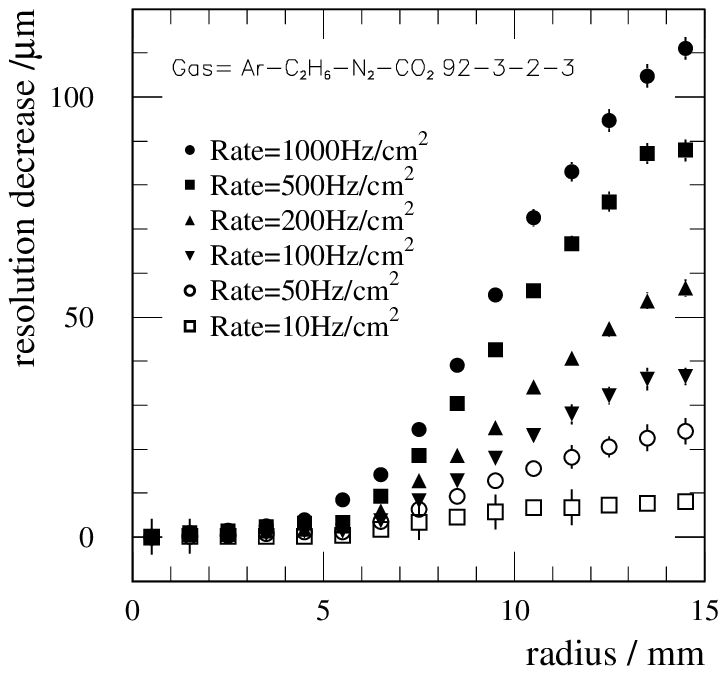,width=\columnwidth}}
\caption{\protect\footnotesize Simulation of the mean position shift
(top) versus radius for different particle rates. Bottom: the
contribution to resolution decrease of the tube. The numbers are of
the same magnitude for rates up to some 100\,Hz/cm$^2$, for higher
rates, the shift increases faster. }
\label{fig=shiftsim} 
\end{figure}
\begin{figure}[t]
\centerline{\psfig{figure=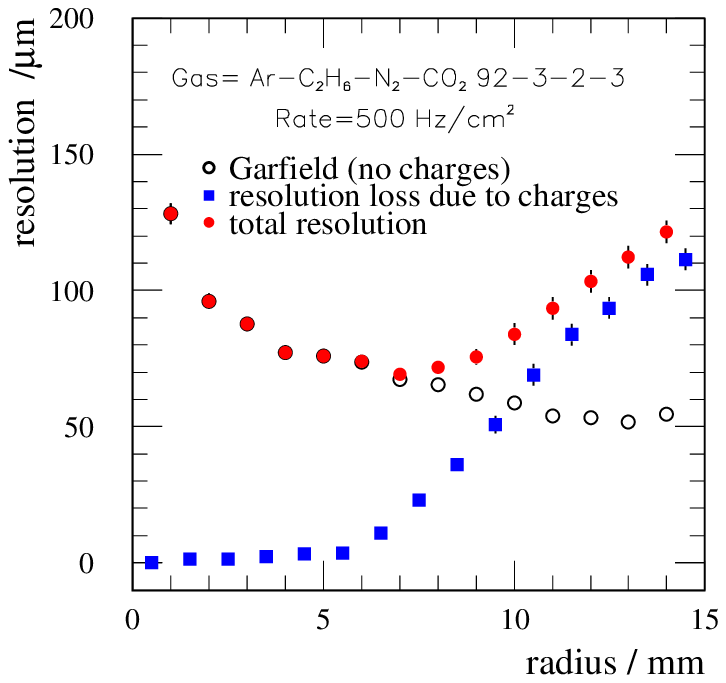,width=\columnwidth}}
\vspace{1mm}
\centerline{\psfig{figure=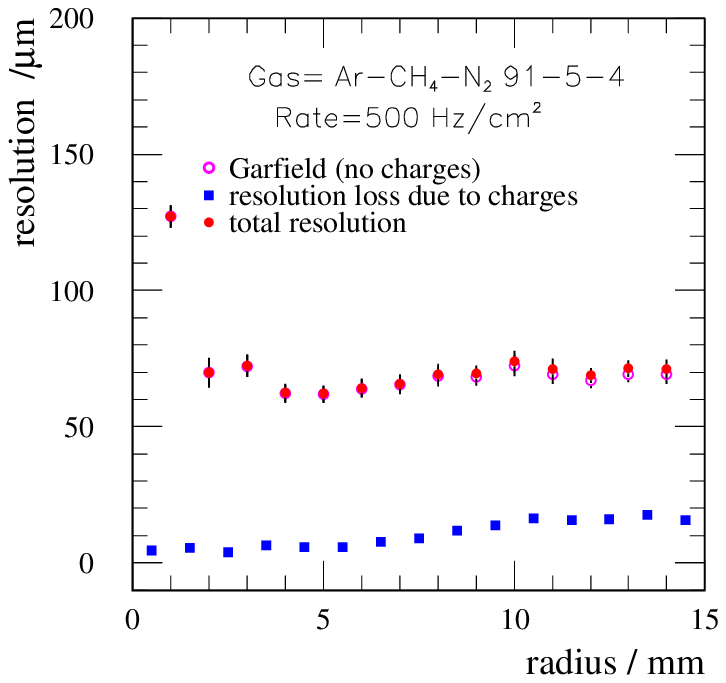,width=\columnwidth}}
\caption{\protect\footnotesize Resolution of the tube. Shown are the
curves coming from a {\protect\sc Garfield} simulation, the 
simulated contribution coming from space charges and the quadratical
sum of both. Top: gas=Ar-C$_2$H$_6$-N$_2$-CO$_2$ 92-3-2-3\,\%, bottom:
gas =  Ar-CH$_4$-N$_2$ 91-5-4\,\%}   
\label{fig=totres} 
\end{figure}
\begin{figure}[t]
\centerline{\psfig{figure=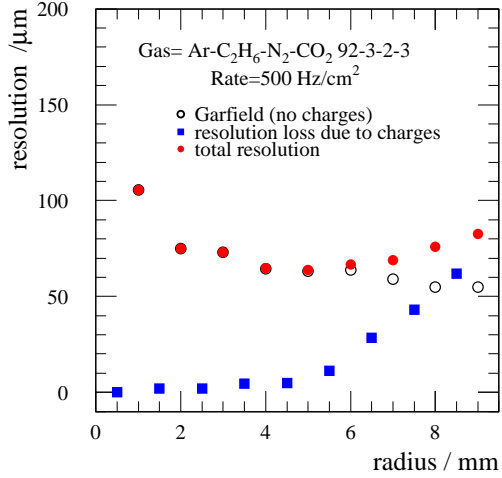,width=\columnwidth}}
\vspace{1mm}
\centerline{\psfig{figure=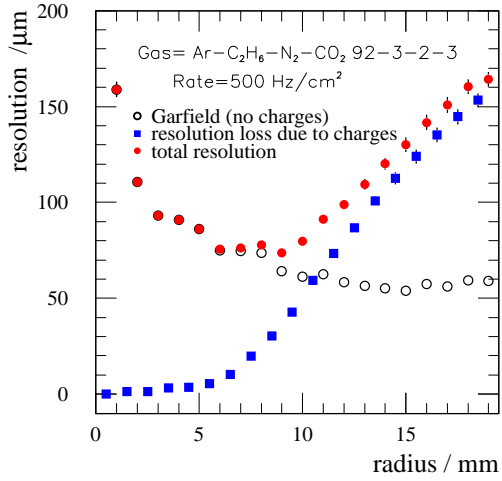,width=\columnwidth}}
\caption{\protect\footnotesize Behaviour of 2\,cm (top) and 4\,cm tubes
(bottom). The corresponding plot for the 3\,cm tube is
fig.\,\ref{fig=totres}(top).} 
\label{fig=totres24} 
\end{figure}
\begin{figure}[t]
\centerline{\psfig{figure=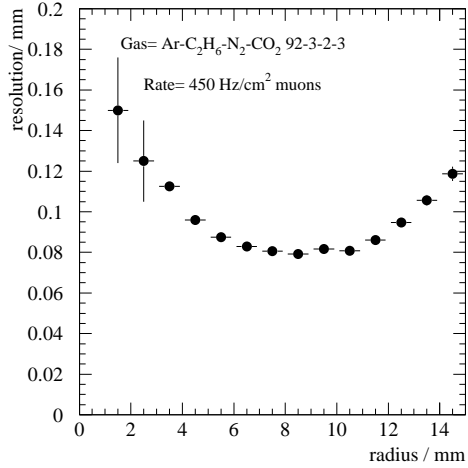,width=0.93\columnwidth}}
\vspace{2.5mm}
\centerline{\psfig{figure=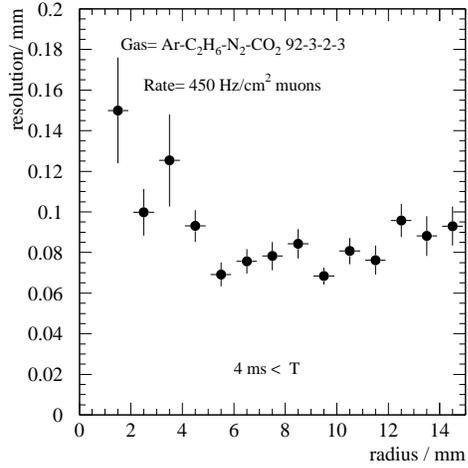,width=0.93\columnwidth}}
\caption{\protect\footnotesize Measured resolution for the gas
Ar-C$_2$H$_6$-N$_2$-CO$_2$ 92-3-2-3\,\%. For the top plot, all events
were taken, for the bottom one only those were selected, where the
previous hit was longer than the ion drift time ago.} 
\label{fig=res129} 
\end{figure}
\begin{figure}[t]
\centerline{\psfig{figure=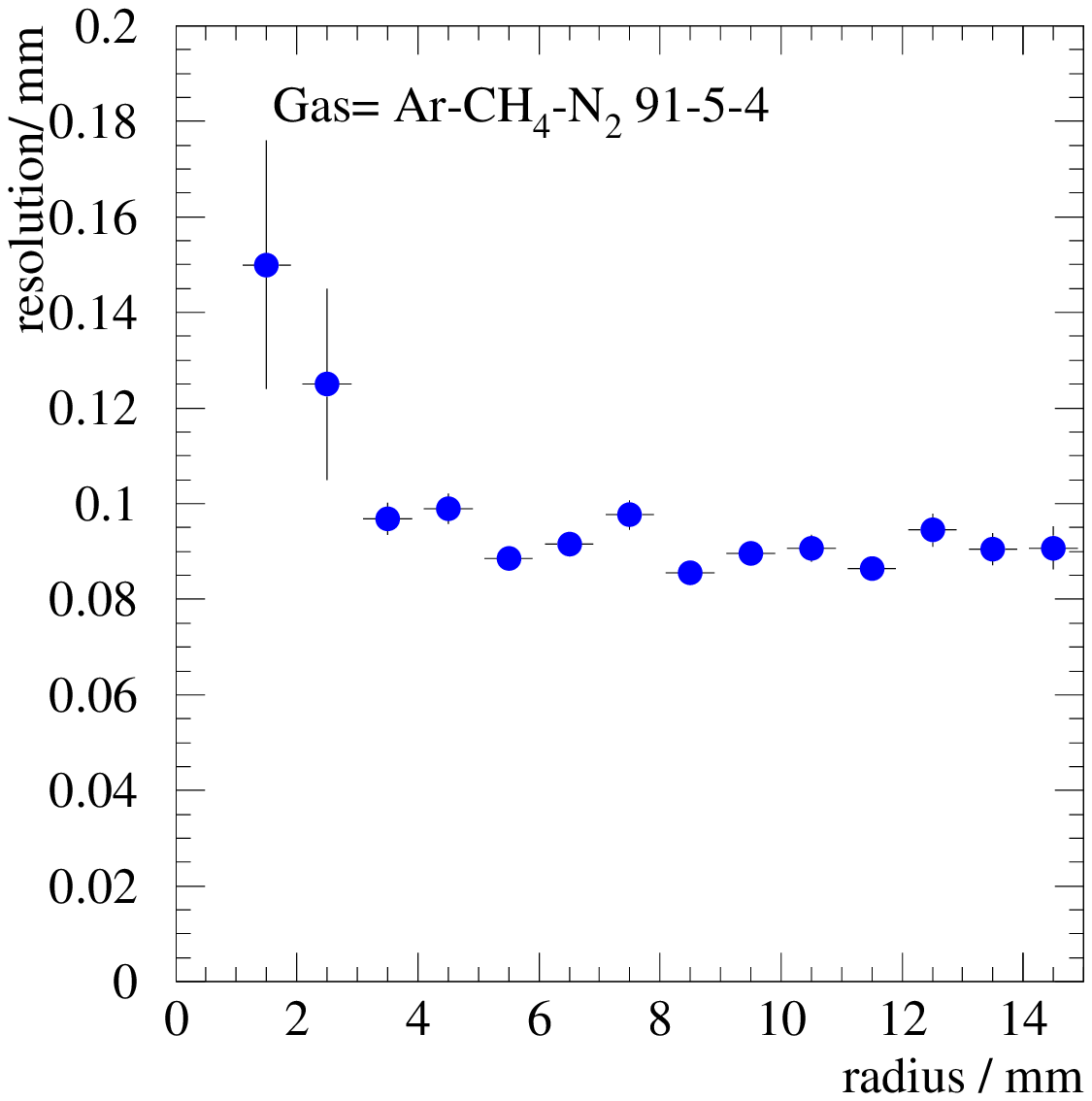,width=\columnwidth}}
\caption{\protect\footnotesize Measured resolution for the gas
Ar-CH$_4$-N$_2$ 91-5-4\,\% with no cut on the history of the tube.
Rate=450\,Hz/cm$^2$.} 
\label{fig=res040} 
\end{figure}
For comparison with the measurements and the analytical calculation
done in the previous section, the mean position shifts for rate=0
... 1\,kHz/cm$^2$ have been calculated (fig.\,\ref{fig=shiftsim},
top). They agree within 5\% with the values shown in
fig. \ref{fig=terr} (bottom) for the gas Ar-C$_2$H$_6$-N$_2$-CO$_2$
92-3-2-3\,\%.\\ 
Fig. \ref{fig=shiftsim} bottom shows the corresponding rms
values of the position shift histogram. This is the contribution to
the resolution loss 
coming from fluctuations. The numbers are non-negligible in the case
of high rates and for high precision drift tubes like those of the
muon spectrometer for {\sc Atlas}. In that case, the resolution of the tubes
is dominated by this effect for a non-linear gas.\\ 
Calculations of drift chamber properties can be done very precisely
using the {\sc Garfield} \cite{garfield} program. It also gives
reliable numbers for the expected resolution (at zero rate).  It was
used to compute 
the resolution of the gases Ar-C$_2$H$_6$-N$_2$-CO$_2$ 92-3-2-3\,\%
and Ar-CH$_4$-N$_2$ 91-5-4\,\% with parameters discriminator
threshold, noise etc set as expected for {\sc Atlas} conditions. The result
is shown in fig. \ref{fig=totres} (open circles). 
On the contrary to the results shown before, the gas gain was doubled to
$4\cdot 10^4$ in order to have the same amount of space charges at the
same rate as expected for the photons of the {\sc Atlas}
background. Also the  measured data that will be shown below refer
to muon data with a gas gain of $4\cdot10^4$.\\  
The results of the
resolution loss for a rate of 500\,Hz/cm$^2$ are shown as
squares. The total resolution (filled circles)  is then given as a
quadratical sum of both contributions. For the first gas, space
charges play an important part for radii $>$ 8\,mm, becoming the
dominant contribution. For the second gas, the resolution is not
affected by space charges even at very high rates. 
The different behaviour of the two gases is in agreement with the
expectations: fig.\,\ref{fig=drift_calc}
already showed that the drift time is nearly uneffected due to space
charges for Ar-CH$_4$-N$_2$ 91-5-4\,\%. For this linear gas, the drift
velocity stays constant with changing electric field - in contrast to
the non-linear gas. \\
An experimental proof of these simulations is shown in
figs. \ref{fig=res129}, \ref{fig=res040}. Tracks have been fitted through the
chamber and the resolution of a single tube was calculated for
different conditions. For the gas Ar-C$_2$H$_6$-N$_2$-CO$_2$
92-3-2-3\,\%, one can see in both, measurement and simulation a
degradation 
of the resolution for big radii in the case of high rates. It 
is not visible if one takes only low rate data or -- as in shown in
fig.\,\ref{fig=res129}(bottom) -- selects from a high rate run only
those events where the previous track was longer than the ion drift
time ago and all space charges are gone.\\
The degradation of the resolution is also not visible for a linear gas
(fig.\,\ref{fig=res040}) where data were taken at the same rate and no
cut on the event selection was applied.\\
The measured resolution is slightly worse (also for low rate) than the
simulation. This is due to the higher noise and a higher threshold for
identification of the pulses needed for our flash ADCs. The simulation
shown refers to TDC measurements as foreseen for {\sc Atlas}. Increasing
the relevant parameters brings the simulation in accordance with our
measurements.\\
Finally, the expected resolution of 2, 3, and 4\,cm tubes shall be
compared using a non-linear gas (for a linear one, the differences are
mostly 
negligible in all cases). The 3\,cm results are shown in
fig.\,\ref{fig=totres}, the others in fig.\,\ref{fig=totres24}.
The 2\,cm tube is nearly unaffected by the high irradiation, a
resolution loss occurs only in the last 2\,mm. On the contrary,  for a
4\,cm tube the resolution is dominated by the space charge
contribution and the mean spatial resolution becomes much worse than
100\,$\mu$m.

\section{Conclusion}

Gain reduction and changes of drift times have been measured as a
function of the rate. For the
gases that have been tested, a good 
agreement  between measurements and calculation was found. Predictions
of the high rate behaviour for new gases can be made without
performing measurements.\\ 
The gas gain reduction due to space charges is similar for most gases
and unavoidable. For a constant rate it could be compensated by
increasing the high voltage.\\
The electric field for drifting electrons is increased because of  space
charges for radii larger than 4\,mm (in a 3\,cm tube). For linear
gases (like Ar-CH$_4$-N$_2$ 91-5-4\,\%)  this may 
have almost no effect on the r-t relation even at very high rates. For
non-linear ones there are already big effects at rates
of 100\,Hz/cm$^2$. Whereas the r-t relation can be adapted for a given
irradiation rate, it is (for practical applications) impossible to
correct for the field fluctuations resulting from the random
distribution  of the times between two hits. This gives a contribution
to the total resolution of the drift tube which for non-linear gases
becomes the dominant limitation of the spatial resolution at large
distances from the wire. 

\section*{Acknowledgements}

We would like to thank our colleagues from the {\sc Atlas} muon
collaboration, especially the people from the test beam group for
their support and many helpful discussions at any time of the day or
night. Special thanks to Rob Veenhof for some very useful ideas. We 
appreciate the hospitality extended by the SMC collaboration
during the taking of these data in their muon beam.  This work has been
supported by the German Bundesministerium f\"ur Bildung,  Wissenschaft,
Forschung und Technologie.   

\end{document}